\documentclass[]{spie}  %>>> use for US letter paper
\usepackage{amsmath,amsfonts,amssymb}
\usepackage{graphicx}
\usepackage[colorlinks=true, allcolors=blue]{hyperref}

\usepackage{caption}
\usepackage{subcaption}

\usepackage{booktabs,caption}
\usepackage[flushleft]{threeparttable}

\usepackage[square,sort,comma,numbers]{natbib}

\title{Design and performance of the PALM-3000 3.5 kHz upgrade}

\author[a,*]{Seth R. Meeker}
\author[a]{Tuan N. Truong}
\author[a]{Jennifer E. Roberts}
\author[a]{J. Chris Shelton}
\author[a]{S. Felipe Fregoso}
\author[b]{Rick S. Burruss}
\author[c]{Richard G. Dekany}
\author[a]{J. Kent Wallace}
\author[b]{John W. Baker}
\author[b]{Carolyn M. Heffner}
\author[a, d]{Dimitri Mawet}
\author[b]{Kevin M. Rykoski}
\author[a]{Jonathan A. Tesch}
\author[a]{Gautam Vasisht}

\affil[a]{Jet Propulsion Laboratory, California Institute of Technology, Pasadena, CA, USA 91109}
\affil[b]{Palomar Observatory, California Institute of Technology, Palomar Mountain, CA, USA 92060}
\affil[c]{Caltech Optical Observatories, California Institute of Technology, Pasadena, CA, USA 91125}
\affil[d]{Department of Astronomy, California Institute of Technology, Pasadena, CA, USA 91125}

\authorinfo{Send correspondence to seth.r.meeker@jpl.nasa.gov. \textcopyright\ 2020. All rights reserved.}

\newcommand{\p}{P3K-II}
\newcommand{\ocam}{OCAM$^{2}$K}
\newcommand{\dc}{$^{\circ}$C}
\newcommand{\e}{e$^{-}$}

\begin{document} 

\maketitle

\begin{abstract}
PALM-3000 (P3K), the second-generation adaptive optics (AO) instrument for the 5.1 meter Hale telescope at Palomar Observatory, was released as a facility class instrument in October 2011 and has since been used on-sky for over 600 nights as a workhorse science instrument and testbed for coronagraph and detector development. In late 2019 P3K underwent a significant upgrade to its wavefront sensor (WFS) arm and real-time control (RTC) system to reinforce its position as a state-of-the-art AO facility and extend its faint-end capability for high-resolution imaging and precision radial velocity follow-up of Kepler and TESS targets. The main features of this upgrade include an EM-CCD WFS camera capable of 3.5 kHz framerates, and an advanced Digital Signal Processor (DSP) based RTC system to replace the aging GPU based system. Similar to the pre-upgrade system, the Shack-Hartmann wavefront sensor supports multiple pupil sampling modes using a motorized lenslet stage. The default sampling mode with 64~$\times$~64 subapertures has been re-commissioned on-sky in late 2019, with a successful return to science observations in November 2019. In 64$\times$ mode the upgraded system is already achieving K-band Strehl ratios up to 85\% on-sky and can lock on natural guide stars as faint as mV=16. A 16~$\times$~16 subaperture mode is scheduled for on-sky commissioning in Fall 2020 and will extend the system’s faint limit even further. Here we present the design and on-sky re-commissioning results of the upgraded system, dubbed \p.
\end{abstract}

% Include a list of up to six keywords after the abstract
\keywords{Adaptive Optics, Shack-Hartmann Wavefront Sensor, Electron Multiplying CCD, Real-time Control,  Digital Signal Processors}

\section{Introduction}
\label{sect:intro}  % \label{} allows reference to this section
The PALM-3000 (P3K) facility adaptive optics (AO) system \cite{Dekany2013} is an "extreme" AO (XAO) upgrade to the previous Palomar AO system \citep[][PALAO]{Troy2000}, engineered to achieve RMS residual wavefront errors down to 105 nm on bright natural guide stars for high-contrast observations, while also supporting general natural guide star use to V$\approx$17. P3K re-purposed the original 241 active actuator deformable mirror (DM) of PALAO as the low-order DM (LODM) of the system, while adding the (then) largest format astronomical DM to date \cite{Roberts2010} with 66 $\times$ 66 PMN actuators (3388 active) as the high-order DM (HODM) in a woofer-tweeter configuration. This high-order correction was necessary for the driving science cases of the system, including imaging of faint Kuiper Belt Objects, stellar evolution studies of pre-main sequence binaries at visible wavelengths, and, especially, the direct imaging and spectroscopy of circumstellar environments and hot, young exo-Jupiters with the P1640 high-contrast integral field spectrograph (IFS) \cite{Hinkley2011}. In addition to the main P1640 survey and the continued utilization of the facility Palomar High Angular Resolution Observer \citep[][PHARO]{Hayward2001}, P3K has served as a vital testbed for other high-resolution and high-contrast instruments and technology demonstrations, including the ten-milliarcseconds per-pixel EM-CCD camera (TMAS) and a myriad of experiments \cite{Galicher2019, Meeker2018, Kuhn2018, Bottom2017} with the Stellar Double Coronagraph \cite{Bottom2016}.

\begin{figure}[t]
\begin{center}
\includegraphics[width=0.7\linewidth]{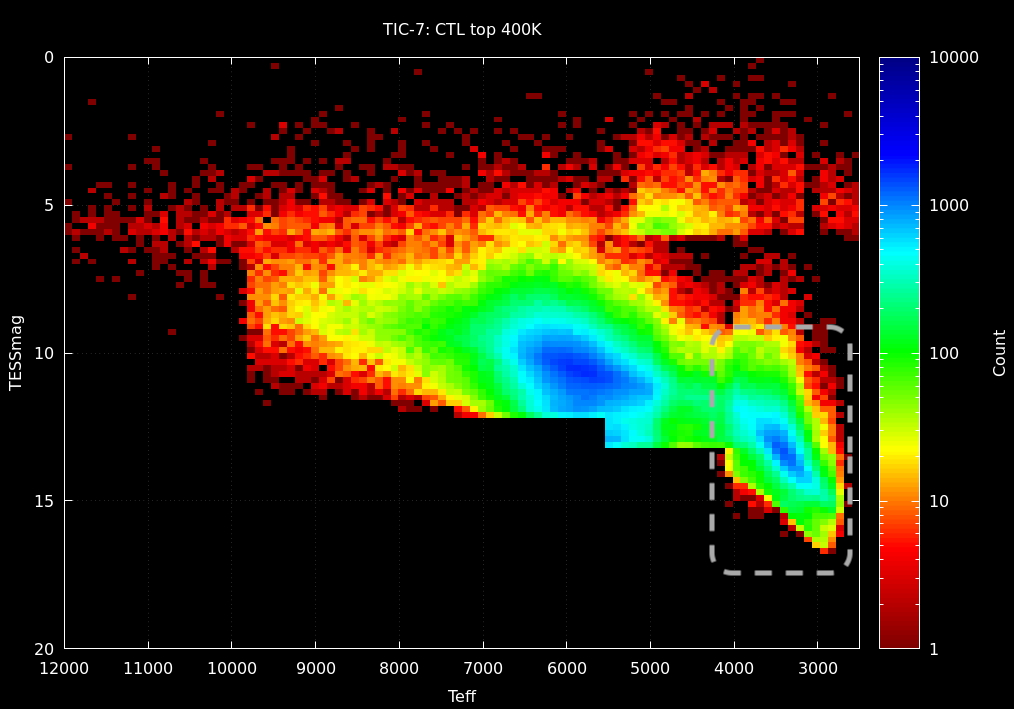}
\end{center}
%\vspace{-12pt}
\caption{Distribution of TESS candidate targets from the TESS Input Catalog v7, as a function of effective temperature, and TESS magnitude. PHARO and PARVI follow-up of a significant fraction of the coolest targets requires NGS operation down to V$\approx$17. Image generated from the visualization tool found at tess.mit.edu/science/tess-input-catalogue/.} \label{fig:tess}
%\vspace{-5pt}
\end{figure}

Following P3K commissioning, the system has undergone continuous improvement and optimization \cite{Burruss2014}, including the roll-out of  32$\times$ and 8$\times$ pupil sub-sampling modes in the wavefront sensor (WFS) to trade high-order correction for decreased measurement noise, thus extending the faint-end performance of the system. This capability enabled and enhanced other science cases for the facility, including monitoring of Pluto and Charon near opposition \cite{Buratti2019} and follow-up of Kepler, K2, and TESS targets to check for stellar multiplicity or background sources. Additionally, after the P1640 survey was completed in 2018, the P1640 dewar and optics were re-purposed to build the Palomar Radial Velocity Instrument (PARVI, Beichman \textit{et al.} in prep), a diffraction limited fiber-fed spectrograph for precision RV measurements. As such, while the SDC continues to operate for high-contrast experiments at Palomar, the dominant science cases for P3K have shifted from high-contrast to high-Strehl applications, with considerable interest in the follow-up of the faintest, reddest targets expected from TESS (see Figure \ref{fig:tess}). This provided the scientific motivation behind an upgrade of the existing CCD50 frame-transfer-CCD WFS camera to an \ocam\ electron-multiplying CCD (EMCCD) with enhanced red-sensitivity from First-Light Imaging, which would simultaneously improve performance and stability on faint targets with lower read-noise, while allowing a faster maximum frame-rate on bright targets.

In addition to the above science motivations, the P3K real-time control (RTC) hardware was beginning to show signs of aging with several of the GPUs used for the wavefront reconstruction calculation failing and being replaced over the years. Beyond that maintenance issue, the RTC hardware and software would require some re-working to support the faster max frame-rate of \ocam. Leveraging our recent experience developing the RTC for NASA's Laser Communication Relay Demonstration (LCRD) integrated optical system \citep[][IOS]{Roberts2018}, a full update of the P3K RTC hardware and software was chosen, based on digital signal processors (DSPs) rather than GPUs, which enabled direct memory access (DMA) transfer from the camera framegrabber for the lowest possible compute latencies.

Following development throughout 2017 and 2018, the upgraded system, referred to hereafter as \p, underwent in-lab integration during Summer 2019, with several nights of on-sky re-commissioning for the default 64$\times$ WFS sampling mode in September through November 2019, and a successful return to science operations in November 2019. The 16$\times$ WFS sampling mode is currently undergoing on-sky commissioning after being postponed due to the COVID-19 facility shutdown. In this proceeding we will present an overview of the \p\ upgrade, performance estimates, and first on-sky results for 64$\times$ mode.

\section{Wavefront Sensor Upgrade}
\label{sect:design}

Details of the  PALM-3000 Shack-Hartmann wavefront sensor and RTC can be found in Dekany et al.\cite{Dekany2013}, but we will briefly summarize the pre-upgrade system here for purpose of camera comparison. The original camera was a 128~$\times$~128 pixel split frame transfer CCD50 from E2V, packaged in a "Li'l Joe" controller head from SciMeasure Analytical Systems. The data are read out at a handful of selectable pixel rates, with corresponding read-noise values. Continuous frame rate selection was enabled by programming a delay between read-out of the final pixel and execution of frame transfer. Camera data was transferred to the Cassegrain cage electronics rack via two cable bundles, where the WFS controller electronics box was mounted, then subsequently sent by CameraLink over optical fiber and optically split to the bank of 8 PCs hosting 16 GPUs that performed wavefront reconstruction.

The new \ocam\ WFS significantly simplifies this configuration. All control electronics exist within the \ocam\ itself with commanding over serial connection through the use of full CameraLink. An EDT full CameraLink transceiver is installed within the \p\ enclosure, allowing us to command \ocam\ and readout frames via a direct fiber link from the \p\ enclosure to the Palomar computer room, reducing weight and heat load in the Cassegrain cage electronics. The \p\ \ocam\ is programmed with a custom firmware from First-Light Imaging to provide a region-of-interest (ROI) readout mode, capable of 3.5 kHz frame rate, with 128 $\times$ 128 pixel format to match the format of our previous camera. Matching the custom \ocam\ ROI to the CCD50 array format is one of several measures taken during this upgrade to avoid modifications to other software, algorithms, and calibration scripts.

\begin{figure}
\centering
\begin{subfigure}[t]{0.47\textwidth}
\includegraphics[width=\linewidth]{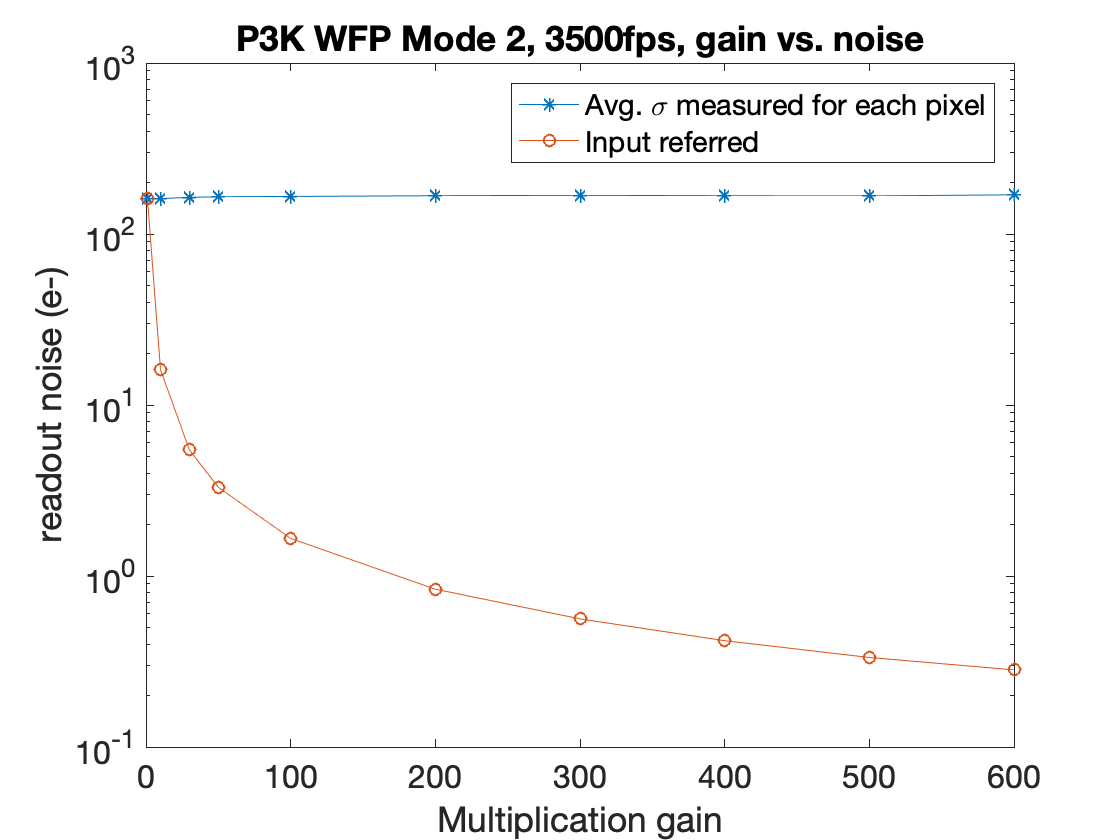}
%\vspace{-5pt}
\end{subfigure}
\vspace{8pt}
\begin{subfigure}[t]{0.47\textwidth}
\includegraphics[width=\linewidth]{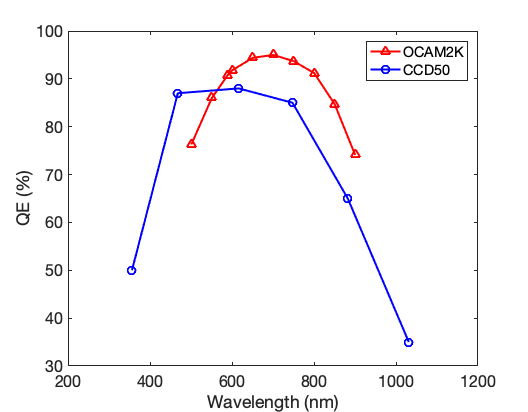}
%\vspace{-12pt}
\end{subfigure}
\caption{(Left) Measurement of the input referenced \ocam\ read-noise, confirming 0.3 e$^{-}$ read-noise at an EM gain of 600. (Right) Manufacturer supplied QE values for CCD50 and \ocam\ detectors.} \label{fig:cams}
%\vspace{-5pt}
\end{figure}

Of course, the other major benefits of upgrading to \ocam\ are the sub-electron read-noise when operating at high EM gain, and the increased red sensitivity. We characterized \ocam\ in the AO lab at Palomar and confirmed the expected 0.3 e$^{-}$ mean read-noise in our ROI when operating at the highest EM gain of 600. The full noise curve is shown in Figure \ref{fig:cams} along with quantum efficiency (QE) curves for CCD50 and \ocam\ that illustrate the increased red sensitivity of \ocam\ at the expense of lowered blue sensitivity. A comparison of relevant detector parameters is tabulated in Table \ref{tab:cams}.

% OCAM Read-noise figure

% CCD50 vs OCAM
\begin{table}[ht]
\begin{threeparttable}
\caption{CCD50 vs. \ocam\ parameters} 
\label{tab:cams}
\begin{center}       
\begin{tabular}{l c c} %% this creates two columns
%% |l|l| to left justify each column entry
%% |c|c| to center each column entry
%% use of \rule[]{}{} below opens up each row
\toprule
\rule[-1ex]{0pt}{3.5ex} Parameter & CCD50 & \ocam \\
\midrule
\rule[-1ex]{0pt}{3.5ex}  Detector Type  & E2V CCD50 Frame Transfer CCD & E2V CCD 220 EMCCD  \\
\rule[-1ex]{0pt}{3.5ex}  Detector Operating Temp & Ambient & -45\dc \\
\rule[-1ex]{0pt}{3.5ex}  Pixel Size & 24 $\mu$m & 24 $\mu$m \\
\rule[-1ex]{0pt}{3.5ex}  Detector Format & 128 $\times$ 128 & 128 $\times$ 128 (custom ROI mode) \\
\rule[-1ex]{0pt}{3.5ex}  Maximum Framerate & 2000 Hz & 3500 Hz \\
\rule[-1ex]{0pt}{3.5ex}  Dark Current  & 150-780 \e pixel$^{-1}$s$^{-1}$ (5\dc) & 15 \e pixel$^{-1}$s$^{-1}$ (-45\dc, EM gain 600) \\
\midrule
\rule[-1ex]{0pt}{3.5ex}  Read-noise$^{*}$  & 9.3 \e (2000 fps) & $<$1 \e \\
\rule[-1ex]{0pt}{3.5ex}             & 7.3 \e (1200 fps)&  \\
\rule[-1ex]{0pt}{3.5ex}             & 3.9 \e (500 fps) &  \\
\rule[-1ex]{0pt}{3.5ex}             & 3.1 \e (200 fps) &  \\
\midrule
\rule[-1ex]{0pt}{3.5ex}  Camera Gain & 0.1 \e/DN (2000 fps) &  30 \e/DN\\
\rule[-1ex]{0pt}{3.5ex}              & 0.4 \e/DN (1200 fps) &  \\
\rule[-1ex]{0pt}{3.5ex}              & 0.7 \e/DN (500 fps) &  \\
\rule[-1ex]{0pt}{3.5ex}              & 2.5 \e/DN (200 fps) &  \\
\bottomrule
\end{tabular}
    \begin{tablenotes}
      \small
      \item $^{*}$See Figure \ref{fig:ocamvccd50} for \ocam\ read-noise as a function of EM gain. Gain used depends on NGS brightness and other conditions. For the vast majority of targets we operate with EM gain $>$ 100, and EM gain 600 for NGS V$\gtrapprox$7.
    \end{tablenotes}
\end{center}
\end{threeparttable}
\end{table} 

\subsection{WFS Optomech}
\begin{figure}[ht]
\centering
\begin{subfigure}[t]{0.65\textwidth}
\centering
\includegraphics[width=\linewidth]{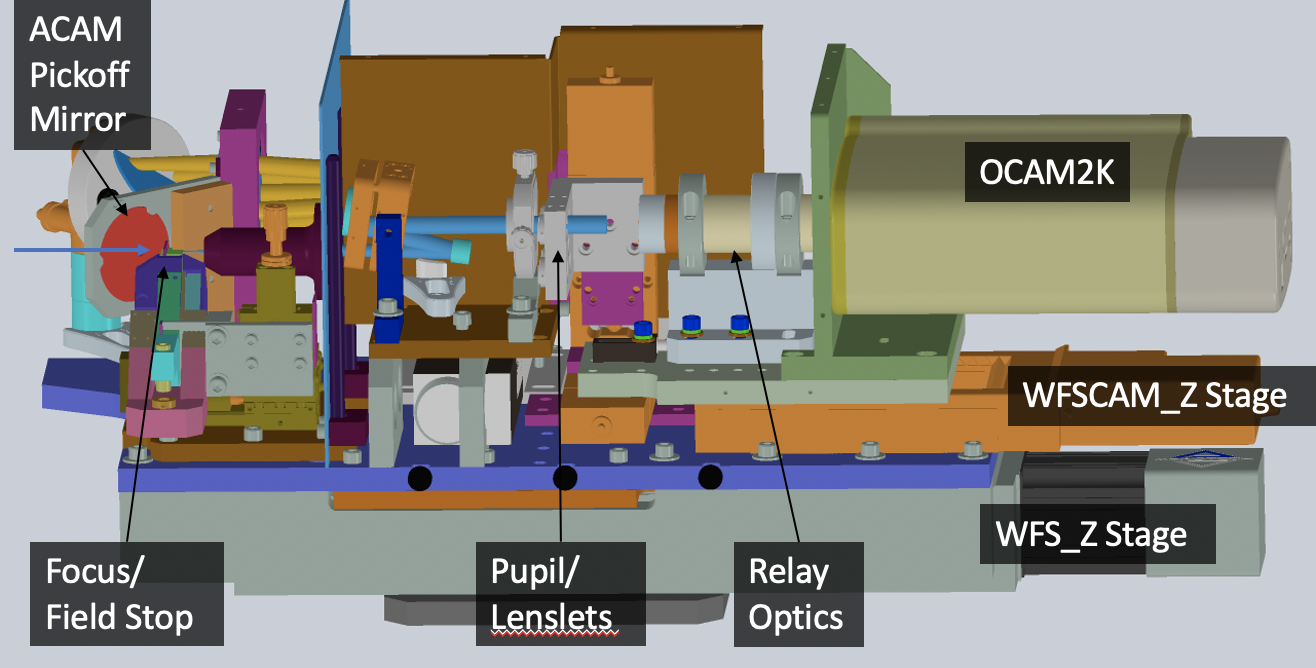} 
\end{subfigure}

\vspace{8pt}
\begin{subfigure}[t]{0.65\textwidth}
\centering
\includegraphics[width=\linewidth]{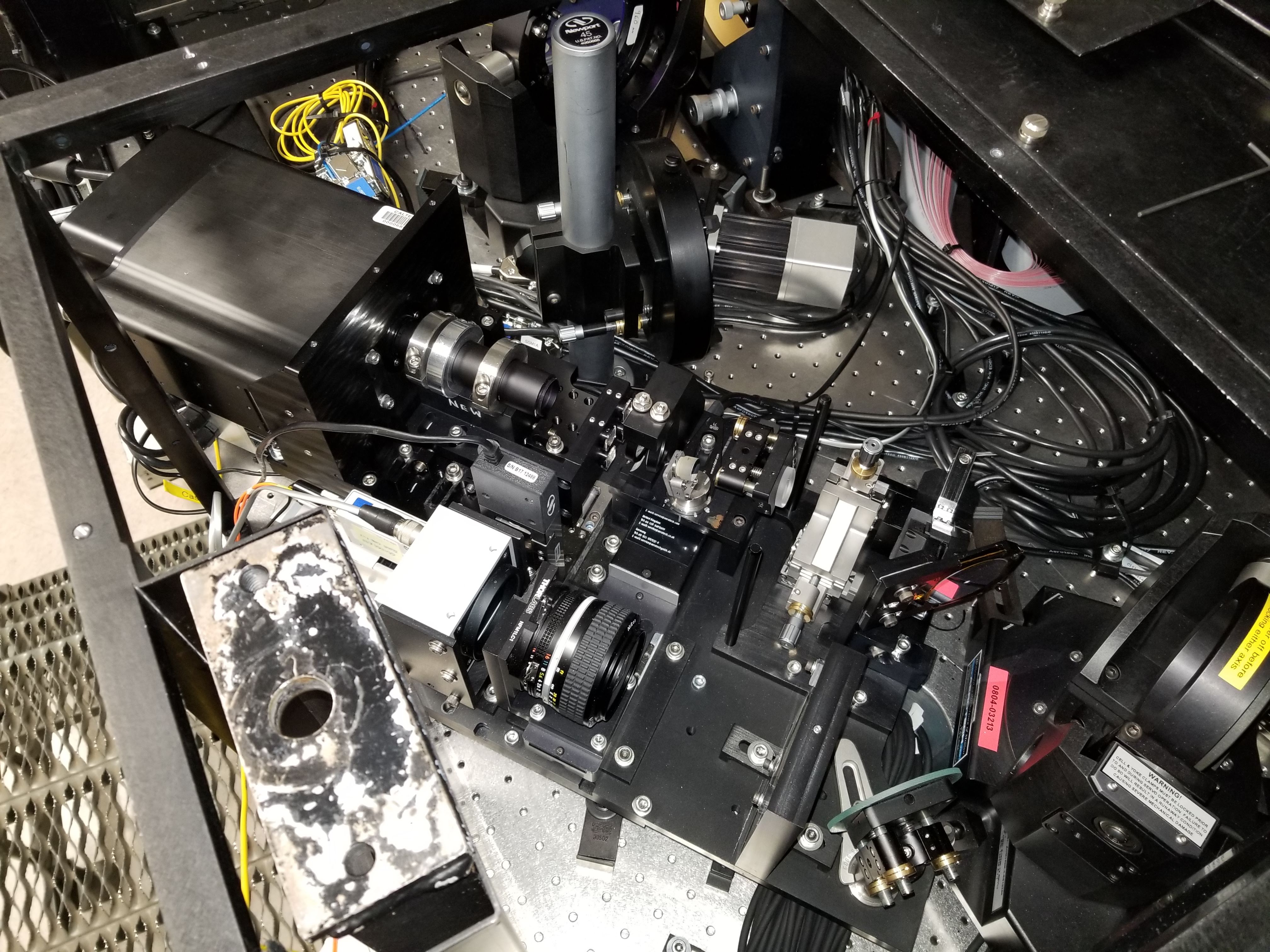} 
\end{subfigure}
\vspace{8pt}
\caption{(Top) Side-view of the \p\ WFS assembly. Light enters from the left, indicated by the blue arrow, passing through an acquisition camera (ACAM) pickoff, field stop, and baffle, then following the blue raytrace to the interchangeable lenslet assembly. Lenslet spots are re-imaged through a two-lens telecentric relay to \ocam. The camera and relay are mounted on a linear stage to adjust for the different focal lengths of the lenslets. (Bottom) a top-down view of the WFS assembly with covers removed for clarity. \ocam\ is in the top left of this image.} \label{fig:ocam}
\end{figure}

Both CCD50 and \ocam\ have 24$\mu$m pixels, which greatly simplified the opto-mechanical interfacing to the new camera. Figure \ref{fig:ocam} shows the CAD model and photograph of the WFS assembly with \ocam\ installed. Because of the identical pixel sizes, the same relay optics could be used for either camera, and thus no optical redesign was required. The baseplate that supports the relay optics and the WFS camera was redesigned with hole patterns and alignment pins to support either CCD50 or \ocam\ to enable the relatively easy swapping of cameras if necessary.
 
% Figure of OCAM in system. Figure of CCD50 in system?

\subsection{Cooling infrastructure}
The \ocam\ detector must be operated at -45\dc, requiring the camera to be liquid cooled. This adds some additional consideration since \p\ is a Cassegrain mounted instrument and is frequently moved between observing floor and AO lab for instrument changes. To reduce the cycling of the quick-release connectors on the \ocam\ camera itself, the flexible hoses from \ocam\ connect to a small assembly of rigid plumbing installed within the \p\ enclosure, including a small expansion tank to accommodate internal pressure swings when moving from the cold observing floor to the warm AO lab. The rigid plumbing exits the enclosure through a bulkhead and terminates with two more quick release connectors. During instrument moves the facility glycol lines are easily attached/detached at this external access point, eliminating extra handling of the enclosure plumbing. As additional safeguards, the enclosure plumbing includes an in-line flow meter, and a length of wetness-sensor rope along the hoses and perimeter of the enclosure. The facility cooling loop and \ocam\ are both shutdown if wetness is detected within the enclosure or if flow is lost to \ocam.

\section{RTC Upgrade}
\label{sect:rtc_design}
The original P3K RTC was based on a GPU implementation of sixteen retail NVIDIA 8800 GTX graphics cards hosted in eight HP Opteron PCs (PC1-8), in conjunction with a Servo PC (PC0) that consolidated the reconstruction results and a Database PC (p3k-telem) that provided overall control and logging of high-speed telemetry. The WFS signal was split using an optical splitter and distributed to PC1-8, each of which captured the frames with EDT framegrabbers. All ten PCs were interconnected using a Quadrics QsNetII 16 port switch for high speed communication. After reconstructed wavefront and computation of mirror commands on the Servo PC, a HODM inter-actuator check was performed in software, as the HODM driver electronics do not include a safeguard against applying too-large interactuator strokes. The TTM, LODM, and HODM commands were then issued through a 4-channel Curtis-Wright FibreExtreme over fiber to the driver electronics in the Cassegrain cage. The HODM control electronics implemented another inter-actuator check in FPGA firmware for redundancy. A detailed explanation of the original P3K RTC architechture and performance can be found in Truong et al.\cite{Truong2012}. Another driving constraint for this upgrade was to maintain the formatting and syntax of all pre-upgrade Database products, commands, and status messages to ensure compatibility with the GUI software, motor automations, and many calibration and analysis scripts.

% RTC block diagram
\begin{figure}[ht]
\centering
\begin{subfigure}[l]{0.47\textwidth}
\hspace{-5pt}
\includegraphics[width=\linewidth]{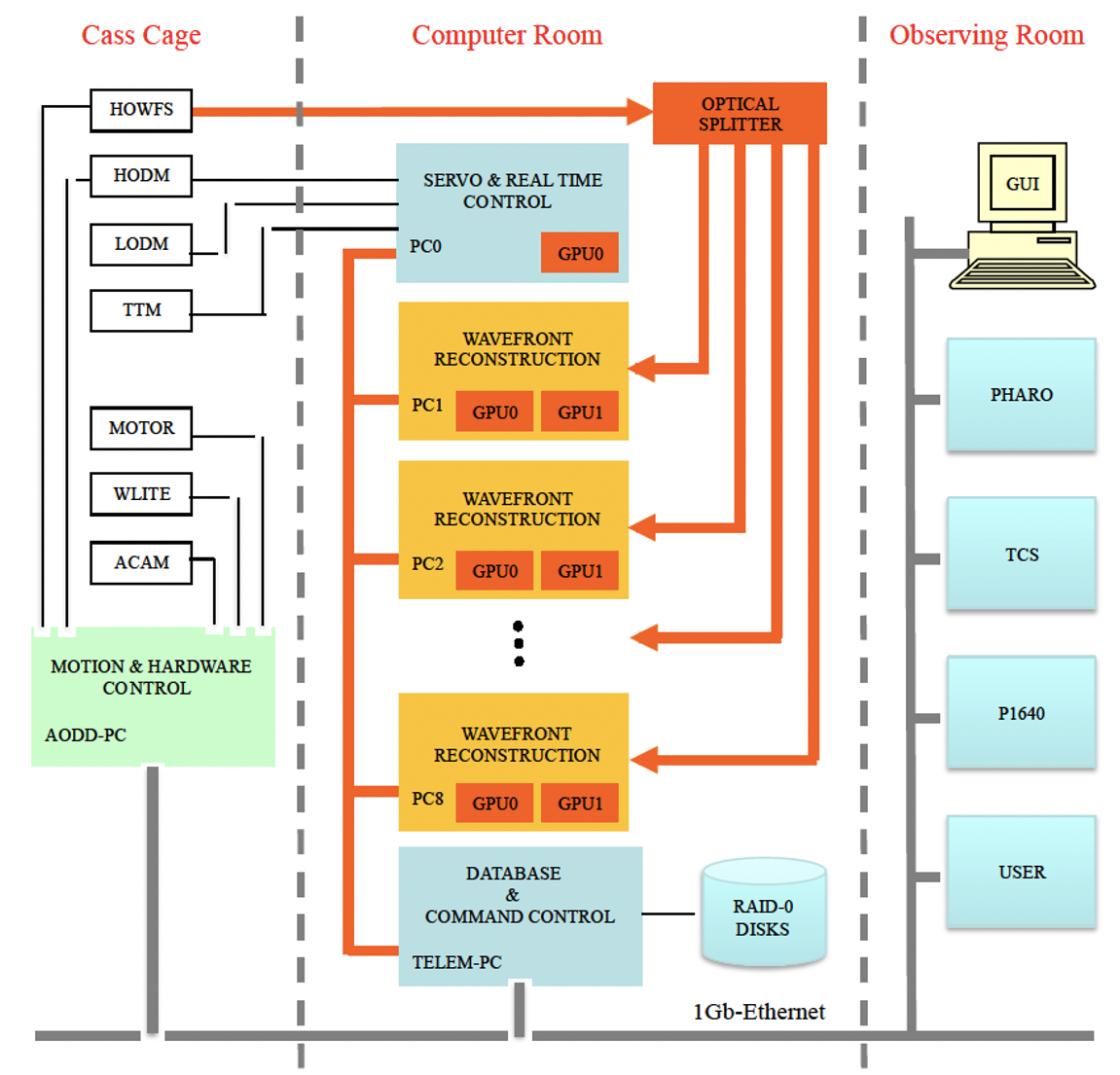}
%\vspace{-5pt}
\end{subfigure}
\vspace{8pt}
\begin{subfigure}[r]{0.47\textwidth}
\hspace{10pt}
\includegraphics[width=\linewidth]{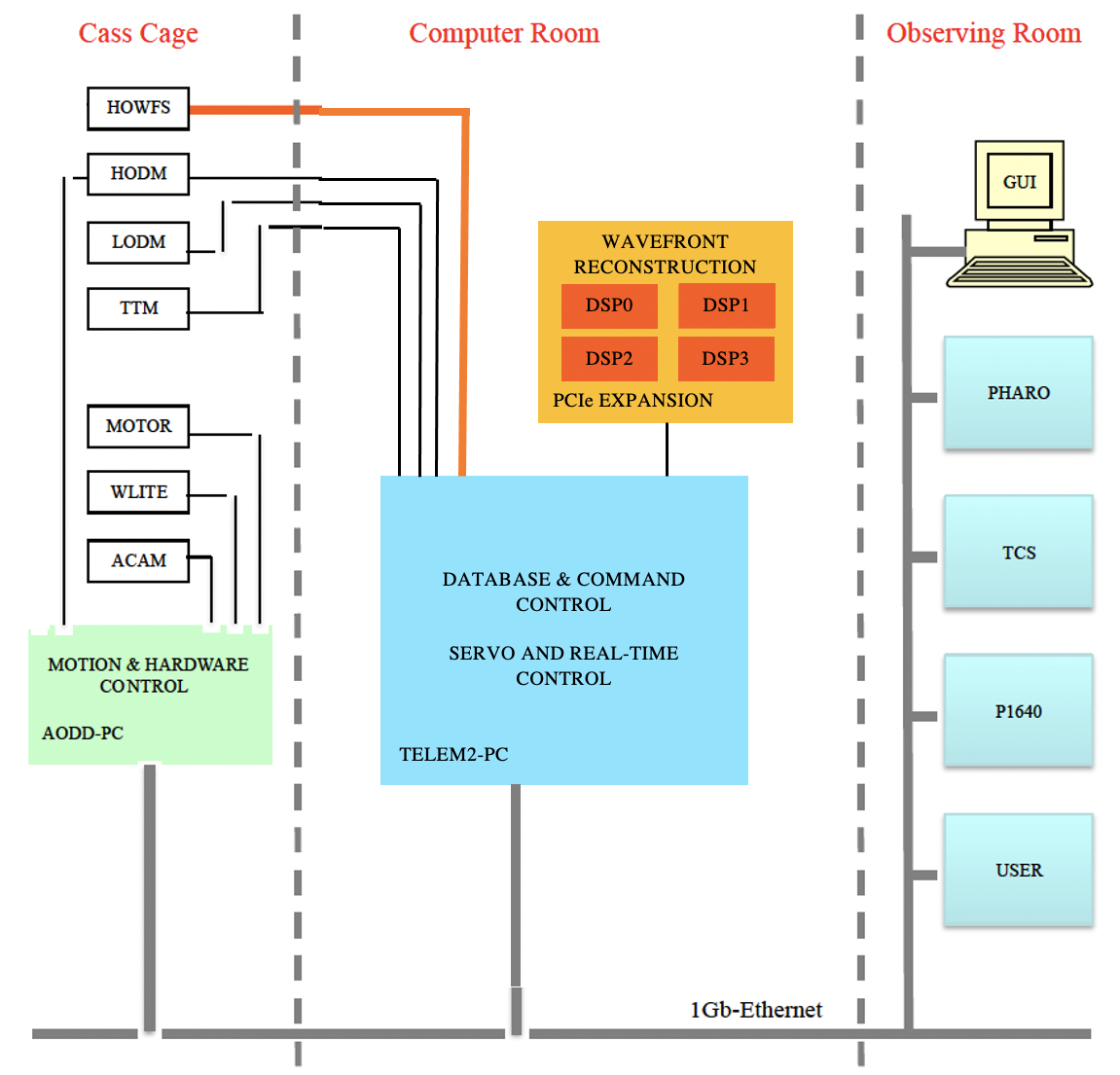}
%\vspace{-12pt}
\end{subfigure}
\caption{(Left) Block diagram of the original P3K RTC hardware configuration. (Right) Block diagram of the \p\ upgrade RTC hardware. The wavefront reconstruction has been consolidated to a single PCIe expansion box housing four DSP boards, connected directly to the main servo and Database PC, eliminating the need for multiple PC boxes and a Quadrics switch.} \label{fig:rtc_block}
%\vspace{-5pt}
\end{figure}

\begin{figure}[ht]
\centering
\begin{subfigure}[l]{0.47\textwidth}
\hspace{-5pt}
\includegraphics[width=\linewidth]{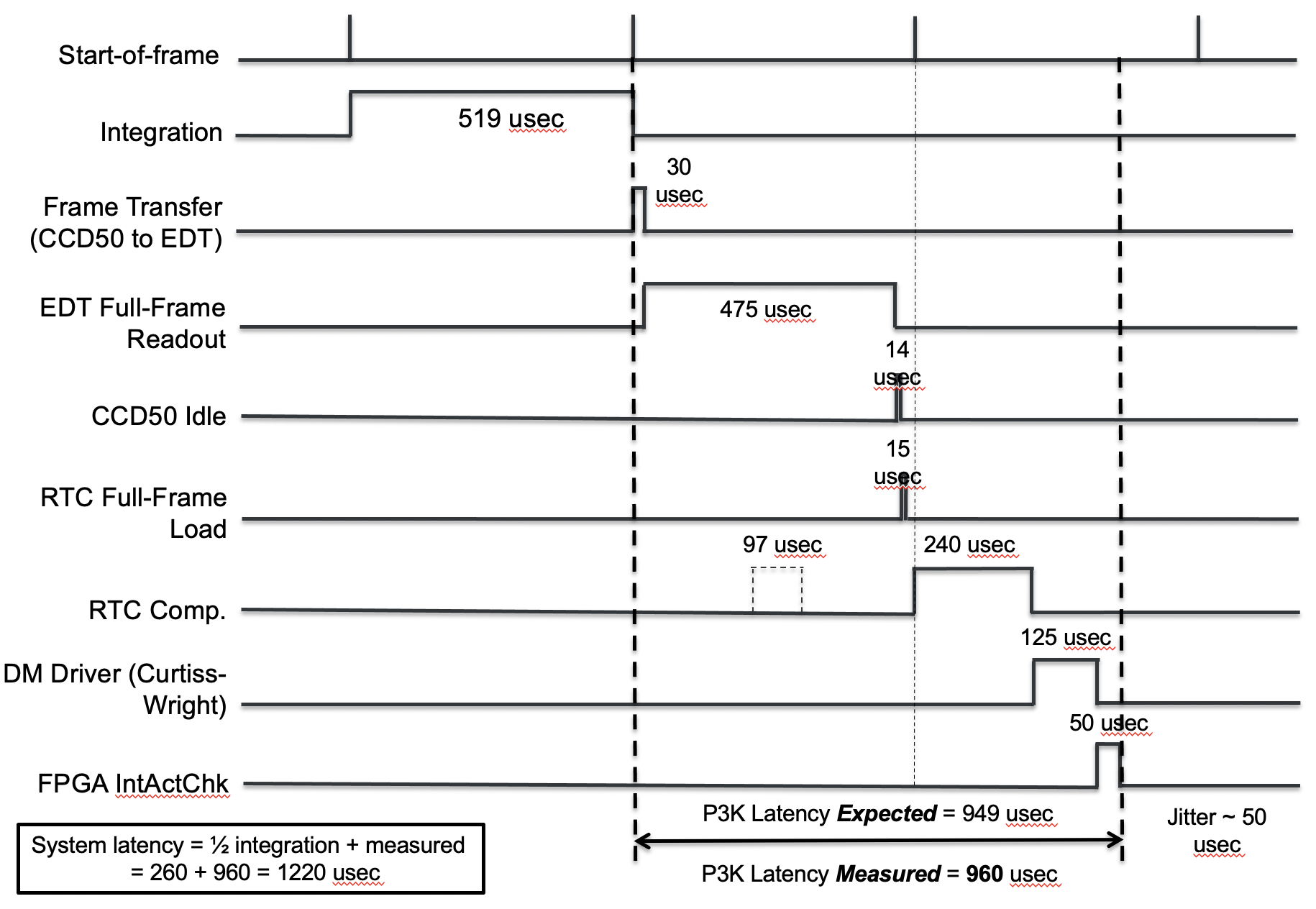}
%\vspace{-5pt}
\end{subfigure}
\vspace{8pt}
\begin{subfigure}[r]{0.47\textwidth}
\hspace{10pt}
\includegraphics[width=\linewidth]{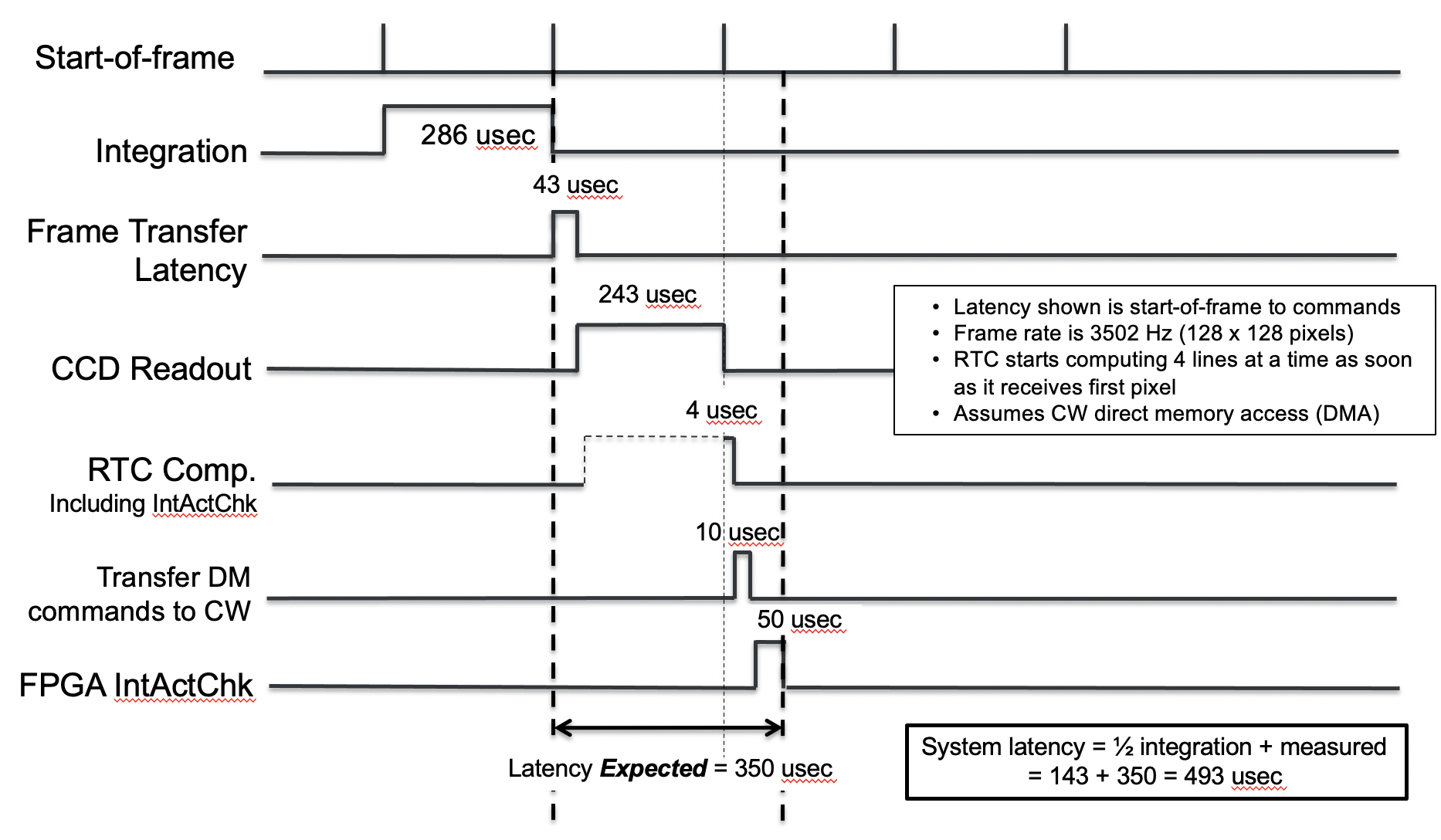}
%\vspace{-12pt}
\end{subfigure}
\caption{Comparison of RTC timing performance for original P3K RTC on the left, and \p\ on the right, from start of frame through completion of the FPGA interactuator check in the HODM control electronics.} \label{fig:rtc_river}
%\vspace{-5pt}
\end{figure}

The \p\ upgraded RTC is based on a digital signal processor (DSP) implementation over four DSPC-8682G2-00A2E DSP cards from Advantech, housed in a PCIe expansion box, which functionally increases the PCIe backplane capacity of the new combined Servo and Database PC (p3k-telem2). WFS frames are captured using an Alacron Fast1703 framegrabber, which was selected because it provides the key innovation of this configuration by enabling direct memory access (DMA) transfer of the camera data to the DSP boards. Camera command and data capture is all performed by a single framegrabber over full CameraLink, removing the need for optically splitting the WFS signal over multiple machines. Instead, we use PCIe multicast functionality to instantly distribute pixel data to all DSPs without executing additional transfers and eliminating any timing jitter that would be incurred by routing through the host machine. Wavefront reconstruction is performed by a vector matrix multiplication, split over the 32 cores of the four DSP boards. Each board receives the full set of WFS pixel data, and each core is responsible for computing a sub-region of the centroids, then multiplying by the associated segment of the reconstructor matrix. After camera integration, the frame is transferred to the optically shielded storage areas for each amplifier region on the chip and readout through the EM registers, with a 43$\mu$s latency from end of integration to first pixel data availability. Computation begins as soon as the first four lines of pixel data are received on the DSPs to parallelize with the collection of pixel data. The full readout time takes 243$\mu$s, and with this parallelization the RTC computation takes only an additional 4 $\mu$s. The HODM interactuator check algorithm is included in this computation stage, before consolidating the HODM position commands. Consolidation of the TTM, LODM, and HODM commands are done on the DSPs, and data is then DMA transferred to a Curtis-Wright board for optical fiber transfer of the commands to the driver electronics. A more detailed description of this DSP RTC implementation is given by Truong et al. (these proceedings). The Cassegrain cage DM electronics are unchanged, and the \p\ servo control block diagram remains the same as that presented in Truong et al.\cite{Truong2012}. The full RTC latency at 3500 Hz from the end of camera integration to sending of DM commands from the Curtis-Wright board takes 300$\mu$s. Since the driver electronics for TTM, LODM, and HODM are unchanged, the latencies associated with execution of those commands after leaving the DSP are unchanged, and are expected to match the timing benchmarked in Dekany et al.\cite{Dekany2013} of 50 $\mu$s for inter-actuator voltage check in FPGA hardware and 60 $\mu$s in the HODM electronics themselves. This suggests a total latency from end of camera integration to HODM movement of 410$\mu$s for the HODM - or 553 $\mu$s (2 frames) total latency when measured from middle of camera integration through HODM voltage settling. Figures \ref{fig:rtc_block} and \ref{fig:rtc_river} provide comparisons between the hardware configuration and timing performance of the original RTC and \p. In-lab direct measurement of this full system latency still needs to be verified.

% RTC rivers

In the DSP RTC implementation for IOS, from which this design was largely inherited, the reconstructor and other calibration table sizes were small enough to store two full copies in dual compute buffers, such that new tables could be loaded during closed-loop operation and activated by simply switching a pointer with no frames skipped in the computation. Part-way through development we discovered that a direct port of the IOS RTC architecture, scaled up for \p\, would not fit on the cache memory resources available in the DSPs. Dual compute buffers would have required a doubling of the number of RTC boards, and possibly an additional PCIe expansion. Instead, when loading a new table, each element in the lone compute buffer is updated as soon as the computation is done with that element. This architecture rework resulted in more efficient resource usage, which ultimately led to some time saved in the RTC computation, at the expense of further development time. Because of the size of the reconstructor table, loading requires several compute cycles. Other tables can still be loaded with sub-frame latency during closed-loop operation, which was especially desirable for using externally computed centroid offsets as a means of injecting external wavefront control. 

\section{Expected AO Performance}
\label{sect:error_budget}
Expected performance is estimated using an error budget tool developed and refined over many years at Caltech and JPL for PALAO, P3K, IOS, and Keck NGAO. Atmospheric assumptions from the error budget, performance and dominant error terms are listed in Table \ref{tab:eb} for a V=7 M3 NGS, comparing \p\ with \ocam\ against the pre-upgrade expectation with CCD50. Considering that the camera and RTC are the only components changed with this upgrade, any error terms related to the rest of the system and the DMs remain unchanged. Aside from some negligible difference in chromatic error due to the different detector QEs, the major difference is the ability to run \p\ faster, decreasing Bandwidth and Measurement error terms. A comparison of performance curves as a function of NGS brightness is included in Figure \ref{fig:ocamvccd50}.

\setlength{\tabcolsep}{24pt}
% CCD50 vs OCAM Error budget comparison
\begin{table}[h!]
\centering
\begin{threeparttable}
\caption{CCD50 vs. \ocam\ AO error budget and atmospheric parameters, assuming a V=7 M3 NGS.}
\label{tab:eb}
\begin{center}       
\begin{tabular}{l c c} %% this creates two columns
%% |l|l| to left justify each column entry
%% |c|c| to center each column entry
%% use of \rule[]{}{} below opens up each row
\toprule
\rule[-1ex]{0pt}{3.5ex}  Atmospheric Parameters  & & \\
\midrule
\rule[-1ex]{0pt}{3.5ex}  Atmospheric $r_{0}$ (500 nm) & \multicolumn{2}{c}{9.2 cm} \\
\rule[-1ex]{0pt}{3.5ex}  Zenith Angle & \multicolumn{2}{c}{30$^{\circ}$} \\
\rule[-1ex]{0pt}{3.5ex}  Ave. Wind Speed & \multicolumn{2}{c}{8 m/s} \\
\rule[-1ex]{0pt}{3.5ex}  Atmospheric $\tau_{0}$ & \multicolumn{2}{c}{3.3 ms} \\
\toprule
\rule[-1ex]{0pt}{3.5ex}  AO settings and Error terms & CCD50 & \ocam \\
\midrule
\rule[-1ex]{0pt}{3.5ex}  Frame rate                 & 990 Hz    & 1760 Hz  \\
\rule[-1ex]{0pt}{3.5ex}  T/T -3dB Bandwidth         & 28 Hz     & 41 Hz \\
\rule[-1ex]{0pt}{3.5ex}  High Order -3dB Bandwidth  & 40 Hz     & 71 Hz \\
\rule[-1ex]{0pt}{3.5ex}  Read-noise                 & 6.8 \e    & 0.3 \e \\
\midrule
\rule[-1ex]{0pt}{3.5ex}  Bandwidth error (nm)                   & 87 & 54 \\
\rule[-1ex]{0pt}{3.5ex}  Measurement error (nm)                 & 79 & 71 \\
\rule[-1ex]{0pt}{3.5ex}  Atmospheric Fitting error (nm)         & \multicolumn{2}{c}{40} \\
\rule[-1ex]{0pt}{3.5ex}  Uncorrectable Instrument Errors (nm)   & \multicolumn{2}{c}{47}\\
\rule[-1ex]{0pt}{3.5ex}  Other high order errors (nm)$^{\dagger}$ & \multicolumn{2}{c}{73} \\
\midrule
\rule[-1ex]{0pt}{3.5ex}  Total high-order wfe (nm)  & 151 & 129 \\
\midrule
\rule[-1ex]{0pt}{3.5ex}  Tip-tilt Error (one-axis)  &   4.5 mas & 4 mas \\
\midrule
\rule[-1ex]{0pt}{3.5ex}  K-band Strehl Ratio        & 82\% & 86\% \\
\bottomrule
\end{tabular}
    \begin{tablenotes}
      \small
      \item $^{\dagger}$Includes: chromatic, dispersion displacement, multispectral, scintillation, calibration, DM stroke, and DM registration error terms.
    \end{tablenotes}
\end{center}
\end{threeparttable}
\end{table} 
\setlength{\tabcolsep}{6pt}

\section{On-sky Commissioning}
\begin{figure}[ht]
\begin{center}
\includegraphics[width=0.75\linewidth]{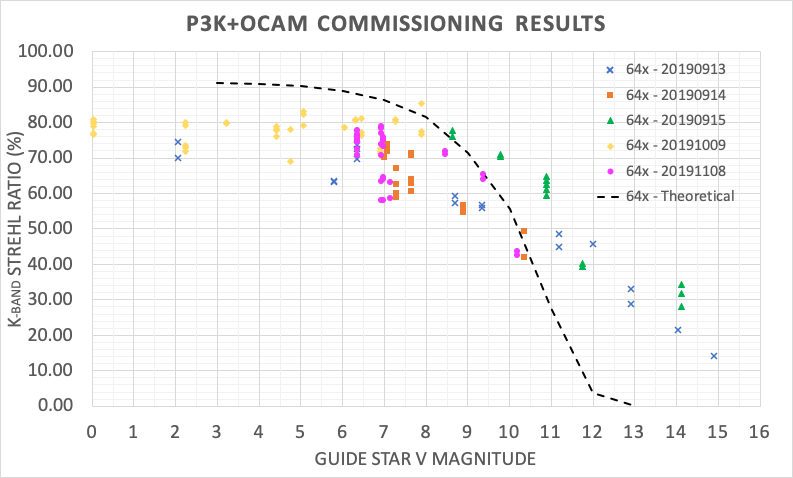}
\end{center}
%\vspace{-12pt}
\caption{K-band Strehl ratio vs. NGS V magnitude throughout the \p\ 64x commissioning phase, broken down by date. Each night's average conditions is included in table \ref{tab:nights}.} \label{fig:ocam_dates}
%\vspace{-5pt}
\end{figure}

\p\ re-commissioning of the 64$\times$ mode took place on the calendar dates of September 12-14, October 08, and November 07, 2019. \p\ is not equipped with an atmospheric turbulence simulator, so just as with the original P3K commissioning, performance as a function of NGS brightness could only be determined on-sky. Data were collected using PHARO, prioritizing K-band observations using K, K$_{s}$, or Brackett-$\gamma$ filters depending on target brightness. Pre-upgrade engineering data was typically collected in these filters, allowing easy comparison to the pre- and post-upgrade system performance, which is assessed by the K-band Strehl ratio
as a function of NGS V magnitude. Over the six re-commissioning nights we collected data across the full span of brightnesses that 64$\times$ mode could safely operate on, with EM-CCD over-illumination potentially occurring for targets brighter than V$\approx$-1. During science operations in November 2019, stable lock was achieved on targets as faint as V=16.2. Figure \ref{fig:ocam_dates} presents all re-commissioning K-band Strehl ratios, divided by observing night, along with the error budget predicted 64$\times$ performance curve. Average conditions from each night are tabulated in Table \ref{tab:nights}. Throughout re-commissioning we primarily targeted M type stars down to V$\approx$11, beyond which spectral type is not known.

% Commiss data by night, forced to be placed along with nightly conditions table

% Nightly conditions table
\begin{table}[ht]
\vspace{20pt}
\centering
\begin{threeparttable}
\caption{Observing conditions from 64$\times$ re-commissioning nights} 
\label{tab:nights}
\begin{center}       
\begin{tabular}{l c c c c c} %% this creates two columns
%% |l|l| to left justify each column entry
%% |c|c| to center each column entry
%% use of \rule[]{}{} below opens up each row
\toprule
\rule[-1ex]{0pt}{3.5ex} UTC Date (YYYYMMDD) & 20190913 & 20190914 & 20190915 & 20191009 & 20191108\\
\midrule
\rule[-1ex]{0pt}{3.5ex}  Ave. Seeing at 500 nm$^{\dagger}$  & 1.45" & 1.42" & 1.51" & 1.04" & 1.35"  \\
\rule[-1ex]{0pt}{3.5ex}  Ave. Wind Direction (Degrees E from N)$^{\dagger\dagger}$ & 87$^{*}$ & 68$^{*}$ & 276$^{*}$  & 207$^{*}$ & 66 \\
\rule[-1ex]{0pt}{3.5ex}  Ave. Wind Speed (m/s) & 1.8$^{*}$ & 2.4$^{*}$ & 3.0$^{*}$ & 3.1$^{*}$ & 2.1  \\
\rule[-1ex]{0pt}{3.5ex}  Ave. Dome Air Temperature & 18.4\dc & 19.1\dc & 20.1\dc & 14.8\dc & 12.7\dc \\
\rule[-1ex]{0pt}{3.5ex}  Ave. $\Delta_{(Primary-Dome) Temp}$ & 1.7\dc & 1.5\dc & 1.4\dc & 0.3\dc & 0.2\dc \\
\bottomrule
\end{tabular}
    \begin{tablenotes}
      \small
      \item $^{\dagger}$Based on FWHM measurements from the Palomar 18-inch dome seeing monitor. Conditions at the 200-inch dome are systematically a bit better than reported by this seeing monitor.
      \item $^{\dagger\dagger}$Direction from which the wind is blowing, not the direction vector of the wind.
      \item $^{*}$The Palomar 200-inch anemometer was not functional on these nights. These data are taken from the Palomar 48-inch site.
    \end{tablenotes}
\end{center}
\end{threeparttable}
\end{table} 

% PSF images
\begin{figure}[ht]
\centering
\begin{subfigure}[l]{0.37\textwidth}
\hspace{-5pt}
\includegraphics[width=\linewidth]{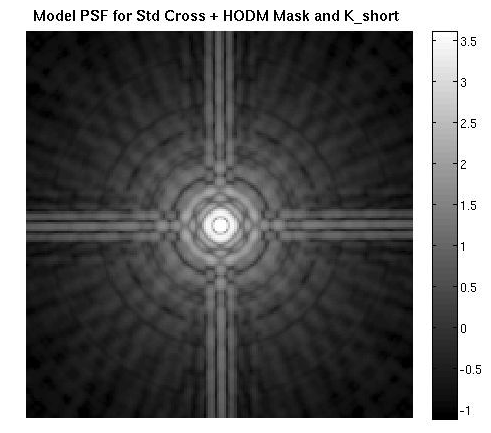}
%\vspace{-5pt}
\end{subfigure}
\vspace{12pt}
\begin{subfigure}[r]{0.37\textwidth}
\hspace{10pt}
\includegraphics[width=\linewidth]{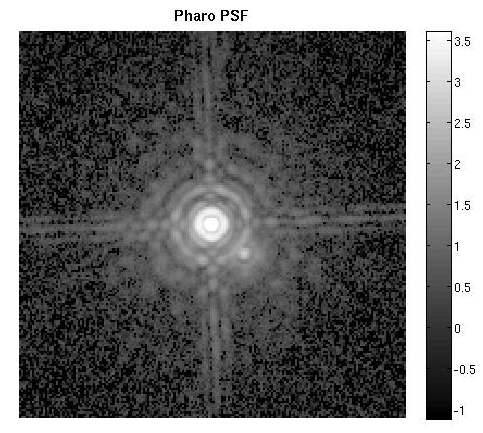}
%\vspace{-12pt}
\end{subfigure}
\caption{Best \p\ on-sky Strehl ratio (85\%) achieved during 64$\times$ re-commissioning - measured through PHARO's K$_{s}$ filter on the evening of UTC 20191009 on the star SAO 90696, class M3III with V=7.88. The model PSF used for normalization to calculate Strehl ratio is shown on the left for reference. An internal ghost from the PHARO 0.1\% neutral density filter is visible in the data at 4~o'clock, and is masked during Strehl computation.} \label{fig:psfs}
%\vspace{-5pt}
\end{figure}

In addition to K-band Strehl ratio vs. NGS magnitude, we characterized the system's on-sky disturbance rejection and -3dB bandwidths for the TTM and both DMs at a variety of gains, checked for any NGS color dependency in the level of correction achieved, and demonstrated closed-loop performance on extended objects by observing Neptune. Initial results from these measurements show the expected performance compared to the pre-upgrade system, but full analysis is on-going. 

% Neptune images
\begin{figure}[ht]
\centering
\begin{subfigure}[l]{0.30\textwidth}
\hspace{-5pt}
\includegraphics[width=\linewidth]{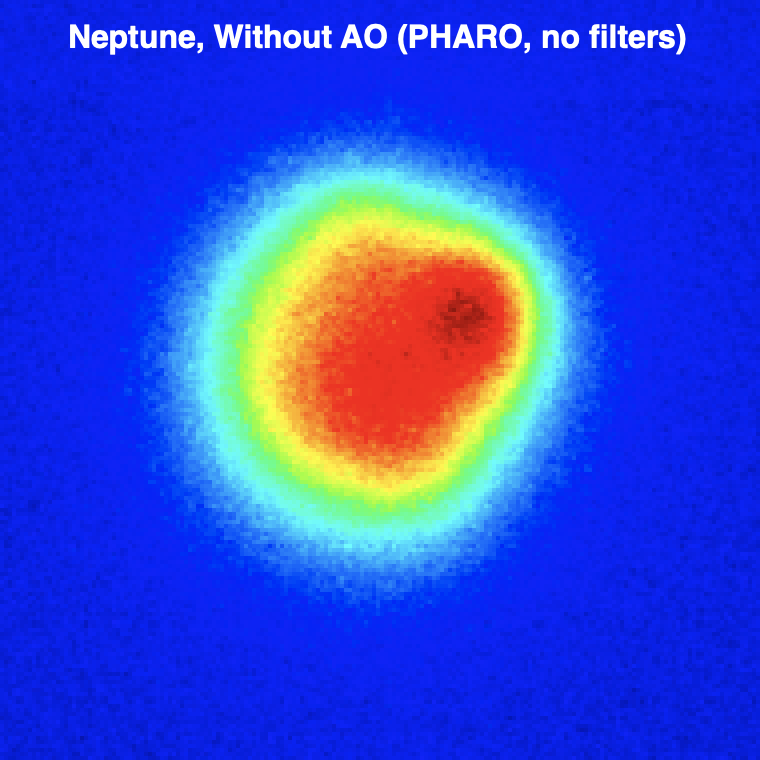}
%\vspace{-5pt}
\end{subfigure}
\vspace{8pt}
\hspace{10pt}
\begin{subfigure}[l]{0.30\textwidth}
\hspace{-5pt}
\includegraphics[width=\linewidth]{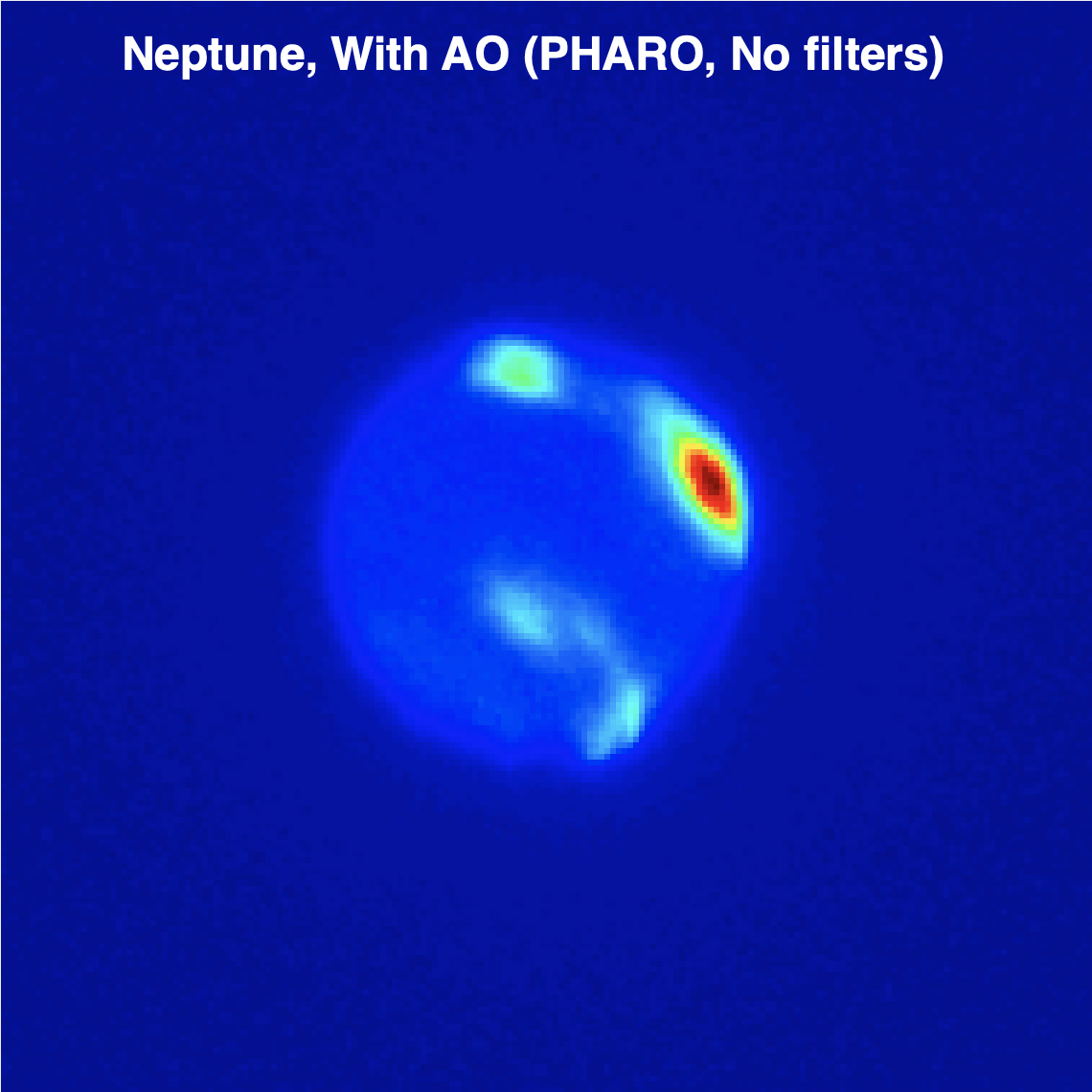}
%\vspace{-5pt}
\end{subfigure}
\vspace{3pt}
\hspace{10pt}
\begin{subfigure}[r]{0.30\textwidth}
\includegraphics[width=\linewidth]{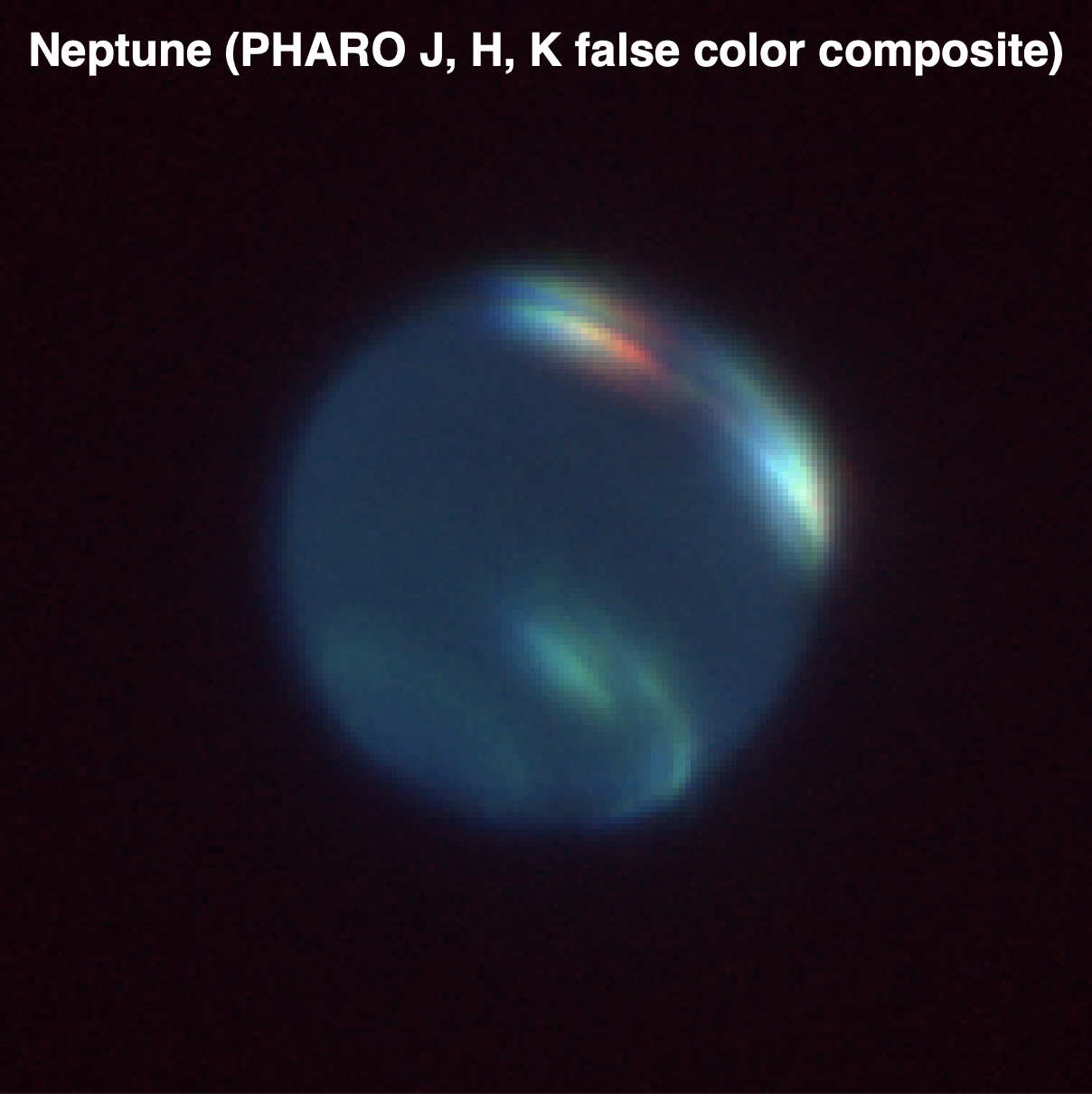}
%\vspace{-12pt}
\end{subfigure}
\caption{(Left) Broadband PHARO images of Neptune collected on UTC 20190915 showing the open versus closed loop resolution. (Right) A 3-color composite of Neptune from the same observations, taken in PHARO J, H, and K filters.} \label{fig:neptune}
%\vspace{-5pt}
\end{figure}

 \begin{figure}[ht]
\begin{center}
\includegraphics[width=0.8\linewidth]{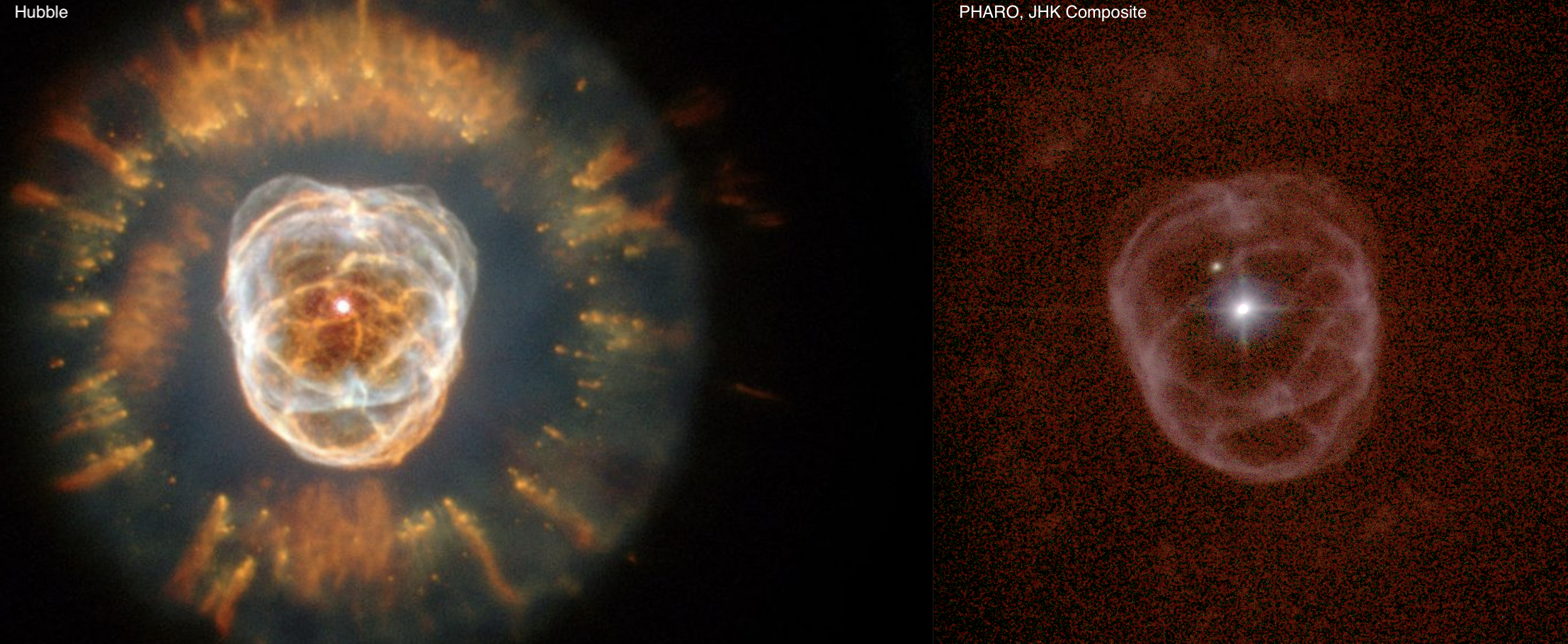}
\end{center}
%\vspace{-12pt}
\caption{PHARO 3-color J, H, K composite image of the planetary nebula NGC2392 during \p\ 64$\times$ re-commissioning, compared side-by-side with Hubble's image on the left. } \label{fig:ocamvccd50}
%\vspace{-5pt}
\end{figure}

\section{Discussion}
\label{sect:discussion}

For comparison with pre-upgrade performance we re-processed and recomputed Strehl ratio for our full archive of P3K+PHARO engineering data covering all WFS sampling modes previously supported. Figure \ref{fig:ocamvccd50} presents this data against \p\ 64$\times$ re-commissioning data, with data series identified by the WFS camera name and sampling mode. For reference, the error budget expected curves are also included for both versions of the system. Overall, we see that the greatly improved faint NGS stability of the \p\ upgrade provides 64$\times$ mode Strehl ratios that rival the pre-upgrade performance over all modes, which depended quite strongly on the prevailing conditions. On the bright NGS end we see \p\ slightly underperforming expectation. The highest 64$\times$ K-band Strehl ratio achieved by \p\ to date is 85\% (Figure \ref{fig:psfs}), versus 91\% pre-upgrade. It is worth mentioning that the best pre-upgrade K-band Strehl ratio was achieved under excellent conditions.  

 \begin{figure}[ht]
\begin{center}
\includegraphics[width=0.75\linewidth]{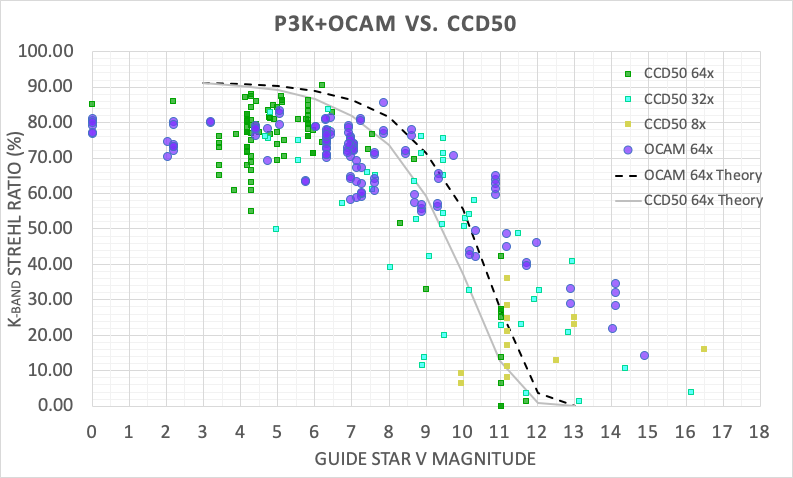}
\end{center}
%\vspace{-12pt}
\caption{K-band Strehl ratio vs. NGS V magnitude throughout the \p\ 64$\times$ commissioning phase compared against pre-upgrade P3K engineering data over all WFS sampling modes (64$\times$, 32$\times$, and 8$\times$). Theoretical error budget curves for \ocam and CCD50 wavefront sensors are included.} \label{fig:ocamvccd50}
%\vspace{-5pt}
\end{figure}

Along those same lines, when examining the re-commissioning performance breakdown by date in Figure \ref{fig:ocam_dates} we see an interesting correlation with the conditions in Table \ref{tab:nights}. Despite having poorer seeing conditions than the two previous nights, consistently better performance was achieved on the evening of 20190915 when the wind shifted direction. This data supports something that has been known anecdotally at Palomar, that better AO performance has historically been achieved when the winds are out of the West. This illustrates further that, even though average seeing and wind speeds may suggest median conditions, a joint analysis of measured Strehl ratios with archived weather data (seeing, temperature, wind at all altitudes, etc.) is needed to truly understand if the system is underperforming relative to previous data.

On faint NGS, \p\ is outperforming the error budget expectation considerably. This was true with the pre-upgrade system as well, but not made as clear due to the larger variation in performance depending on prevailing conditions. This performance boost is  attributed to improvements in the Shack-Hartmann centroiding algorithm\cite{Shelton1997} and a Bayesian reconstructor algorithm\cite{Wild1998} that computes an optimal regularization parameter based on a measurement of $r_{0}$ extracted from the WFS residuals. These improvements were added to the system throughout the original P3K optimization effort to improve faint NGS performance, but are not captured in an analytical way in our error budget. We will instead explore the benefit of these techniques in future work via numerical simulations.

\section{Future Work and Lessons Learned}

This major upgrade was not without some issues and lessons learned along the way. A development issue related to limited DSP memory was already mentioned earlier. Following the 64$\times$ re-commissioning phase, we also discovered significantly higher flexure in the WFS mounting coming from the focus stage supporting \ocam, causing a misalignment of the \ocam\ with respect to the lenslets whenever crossing meridian. After subsequent engineering time on-sky, this issue is well understood, and can be resolved with a calibrated adjustment to the lenslet positions when crossing meridian during an observation or telescope slew. Our investigation suggests the increased pitch and yaw error could result from the stage's roller bearings losing their pre-load. Repair is underway, though the cause of this remains a mystery. We are left to speculate that the issue was present prior to the upgrade, but only revealed itself with the additional weight of the new camera.

The immediate near-term goal for the system is the commissioning of the 16 $\times$ 16 subaperture WFS sampling mode, operating in a Fried geometry with the LODM, which is expected to further extend the faint limit of the system by 2 magnitudes. This work was originally planned for early-mid 2020, but was postponed to late 2020 due to the COVID-19 pandemic. Further analysis is also required to quantify the  improved performance, if any, on redder NGS. Preliminary analysis looking for higher throughput at redder wavelengths has so far shown no appreciable difference, however, our pre-upgrade engineering data predominantly targeted hotter stars and few M stars. In our future \p\ engineering we will plan to target a broader range of spectral types to carry out this comparison. To further reduce the full system latency and improve bright NGS performance, we also hope to incorporate the adaptive control techniques previously demonstrated on P3K and IOS\cite{Tesch2019} as part of the standard \p\ operation, which will require a modest expansion of the RTC hardware.

\section{Conclusion and Future Work}
\label{sect:conclusion}
We have presented the design and first on-sky results from \p, a major upgrade to the PALM-3000 NGS AO system at Palomar. The upgrade to a \ocam\ EMCCD WFS has significantly improved the reliability of the system's default 64 $\times$ 64 subaperture mode, providing a stable loop-lock on-sky on NGS as faint as V=16. On-sky re-commissioning results from this mode show it is already performing well even in unfavorable conditions, though a direct comparison with pre-upgrade performance will require more extensive meta analysis to include weather and environmental factors along with the Strehl ratios achieved. In the very near future we expect to deliver the 16 $\times$ 16 subaperture mode, which will enable diffraction limited imaging follow-up and precision RV followup with PARVI on some of the faintest and coolest planet-hosting candidates identified by TESS.

\subsection* {Acknowledgments}
The authors would like to acknowledge the excellent (and patient) support of the Palomar Observatory staff during this instrument upgrade, re-commissioning, and beyond as we worked to understand this new system. This research was carried out in part at the Jet Propulsion Laboratory, California Institute of Technology, under a contract with the National Aeronautics and Space Administration. 

%%%%% References %%%%%

\bibliography{report}   % bibliography data in report.bib
\bibliographystyle{spiejour}   % makes bibtex use spiejour.bst
 
%%%%% Biographies of authors %%%%%

%\vspace{2ex}\noindent\textbf{Seth Meeker} is an Optical Engineer at the Jet Propulsion Laboratory working in the Adaptive Optics and Astronomical Instrumentation group. He received his BS degree in Astrophysics from the University of California, Los Angeles in 2009, and his M.S. and PhD degrees in Physics from the University of California, Santa Barbara in 2013 and 2017, respectively. His research interests include the development of adaptive optics and superconducting detector technologies for various applications, including exoplanet direct imaging and optical communication.

%\vspace{1ex}
%\noindent Biographies and photographs of the other authors are not available.

%\listoffigures
%\listoftables

\end{document}